\documentclass[sigconf]{acmart} 
\AtBeginDocument{%
  \providecommand\BibTeX{{%
    \normalfont B\kern-0.5em{\scshape i\kern-0.25em b}\kern-0.8em\TeX}}}

\setcopyright{none}
\copyrightyear{2022}
\acmYear{2022}
\acmDOI{XXXXXXX.XXXXXXX}

\acmConference[VRST '22]{ACM Symposium on Virtual Reality Software and Technology}{Nov 29 -- Dec 1,
  2022}{Tsukuba, Japan}
\begin{document}

\title[Eliciting Multimodal Interactions in a Multi-Object Augmented Reality Environment]{Eliciting Multimodal Gesture+Speech Interactions in a Multi-Object Augmented Reality Environment}


\author{Xiaoyan Zhou}
\affiliation{%
  \institution{Colorado State University}
  \streetaddress{}
  \city{Fort Collins}
  \state{Colorado}
  \country{USA}}
\email{Xiaoyan.Zhou@colostate.edu}

\author{Adam S. Williams}
\affiliation{%
  \institution{Colorado State University}
  \streetaddress{}
  \city{Fort Collins}
  \state{Colorado}
  \country{USA}}
\email{AdamWil@colostate.edu}

\author{Francisco R. Ortega}
\affiliation{%
  \institution{Colorado State University}
  \streetaddress{}
  \city{Fort Collins}
  \state{Colorado}
  \country{USA}}
\email{fortega@colostate.edu}

\renewcommand{\shortauthors}{Trovato and Tobin, et al.}

\begin{abstract}
  As augmented reality technology and hardware become more mature and affordable, researchers have been exploring more intuitive and discoverable interaction techniques for immersive environments. In this paper, we investigate multimodal interaction for 3D object manipulation in a multi-object virtual environment. To identify the user-defined gestures, we conducted an elicitation study involving 24 participants for 22 referents with an augmented reality headset. It yielded 528 proposals and generated a winning gesture set with 25 gestures after binning and ranking all gesture proposals. We found that for the same task, the same gesture was preferred for both one and two object manipulation, although both hands were used in the two object scenario. We presented the gestures and speech results, and the differences compared to similar studies in a single object virtual environment. The study also explored the association between speech expressions and gesture stroke during object manipulation, which could improve the recognizer efficiency in augmented reality headsets.
\end{abstract}




\maketitle

\section{Introduction}
Easy-to-remember gestures produce high usability interfaces~\cite{NIE+04}. A gesture set that does not align with users' expectations or mental models often leads to frustrating user experiences~\cite{DEZ+11}. Wobbrock et al. introduced an elicitation methodology to collect proposed gestures from users~\cite{WOB+09}, facilitating intuitive gestures design without implementing perfect recognizers in advance. Prior findings proved that users prefer to choose input modalities based on their needs during the interaction ~\cite{BAI+18, KAR+18, MOR12}. Previous studies have explored gesture set design in different devices and interfaces ~\cite{WOB+09, MOR12, RUI+11, COH+12, VOG+22}, and several have been done with multimodal interactions in AR or VR ~\cite{TVCG, PHA+18}. However, few to no researches have involved multimodal interactions in a multi-object AR or VR environment. These prior works raised multiple questions that we were explored during this study: Does multimodal interaction look different when having multi-object virtual environments? Does a multi-object environment impact the gesture and speech proposals? What gestures do users prefer with multiple object manipulation, and are there any differences from single object manipulation? The raised questions drive our motivation to understand if previous single object studies may transfer to more realistic environments. For this work, an elecitation study was conducted for multimodal interaction in AR with a Wizard of Oz (WoZ) experiment design (i.e., a researcher emulating a live system)~\cite{WOB+09, VAT+22}. It involved 24 participants, 22 referents (i.e., command) in augmented reality (AR), and a head-mounted display (HMD). It yielded 528 proposals, and we generated a winning gesture set with 25 gestures after utilizing binning and ranking. We compared our single virtual object manipulation proposals to the findings from prior studies in a single object virtual environment~\cite{TVCG, ISS, PHA+18}. For multiple object manipulation proposals, we compared them with the proposed gestures of single virtual object manipulation in our study. To the best of our knowledge, this is the first study to conduct multi-object mid-air interaction using optical-see through augmented reality headsets.

\section{Related Work}
Elicitation methodology has been widely used in the HCI field to collect user-defined gestures. Wobbrock et al. popularized an elicitation methodology to collect proposed gestures from users~\cite{WOB+09}, which aims to assist in designing more intuitive~\cite{WOB+09}, guessable~\cite{WOB+05}, learnable, and memorable~\cite{MIG+13} interaction techniques. Morris et al. found that people prefer gestures proposed by end-users, which were less complex than ones designed by human-computer interaction (HCI) experts~\cite{MOR+10}. Based on recent literature review results~\cite{VIL+20}, over two hundred studies have adopted the use of an elicitation methodology in their work. Prior findings proved that users prefer to choose input modalities based on their needs during the interaction, such as choosing gestures over speech in an quiet environment~\cite{BAI+18, KAR+18, MOR12}. Wobbrock et al.~\cite{WOB+09} discovered that having synonyms in a user-defined gesture set can increase the guessability of proposed gestures. A multimodal elicitation study provides the opportunity to create multimodal synonyms~\cite{MOR12}, which can offer users different modalities to achieve the same effect. 

Nevertheless, most elicitation studies involving mid-air gestures in augmented reality (AR) only considered single object manipulation in a single object virtual environment~\cite{PIU+13, PHA+18, TVCG, ISS}. Pham et al. conducted an elicitation study with an AR headset that included a scenario of single building manipulation among multiple buildings~\cite{PHA+18}. However, the whole model was attached to a physical surface so that the elicited gestures in the study were not mid-air gestures. Moreover, as far as we know, no research has been done in multimodal interactions with multiple object manipulations in AR. Piumsomboon et al. implemented an elicitation study in AR (video-see through) that asked participants to select multiple objects and the elicited gestures were surface gestures~\cite{PIU+13}. Wittorf et al. adopted an elicitation methodology for exploring mid-air gestures with a wall display~\cite{WIT+16}. Danielescu and Piorkowski conducted an elicitation study to explore free-space gestures with a projector display that included multiple target selection among a set of photos~\cite{DAN+21}. However, the referent showed that photos were selected one by one, which could bias the participants' gesture proposal. Wobbrock et al. found that users preferred one hand over two hands for tabletop interaction ~\cite{WOB+09}. We were interested in whether users preferred two hands for more than one object manipulation. This study aimed to understand the multimodal interaction in a multi-object virtual environment compared to a single object virtual environment. Furthermore, we were interested in the difference between single object manipulation and multiple object manipulation.

\section{Study Design}

This study conducted the elicitation experiment using a similar process as previous work~\cite{WOB+09,TSA18,VAT+22}. 22 tasks (i.e., referents) were used for each modality during this work. Of those, 17 basic referents were selected based on their inclusion in prior works~\cite{TVCG, ISS}, while the other 5 were developed to be multi-object versions of basic referent. Referents included six translations (along x, y, and z axes), six rotations (around x, y, and z axes), three abstracts (create, destroy, and select) and two scales (enlarge and shrink). For multiple object manipulation, only abstract and scale referents were included. There were three experiment blocks in this study, which included modality gesture only (G), speech only (S), and gesture plus speech (GS). Each block took approximately 10 minutes, plus two questionnaires and three surveys. The experiment lasted approximately 45 minutes.

\subsection{Participants} 

The study involved 24 participants (12 female, 12 male). Due to the pandemic, it was difficult to recruit outside of the Computer Science (CS) department, therefore 17 out of 24 participants came from CS. Their ages ranged from 18-34 years (Mean = 23.42, SD = 4.20). All participants had previously used multi-touch devices, nineteen had used motion sense devices (e.g. Xbox Kinect or Nintendo Wii Motion), sixteen had used virtual reality headsets, and three had used augmented reality headsets.

\subsection{Setup}

\begin{figure}
  \begin{minipage}[t]{0.5\linewidth}
    \centering
    \includegraphics[width=0.89\textwidth]{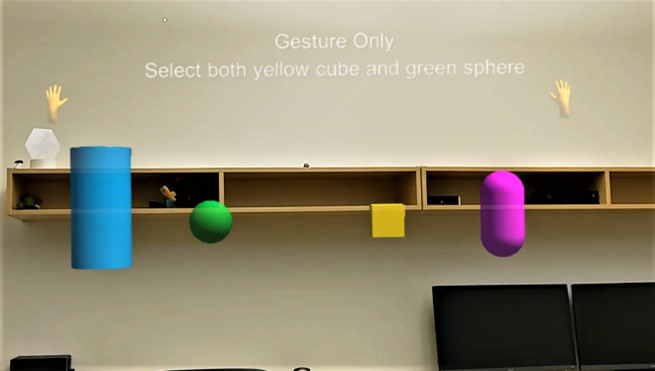}
    \caption{Participant view}
    \Description{The view presented to participants in the experiment. Participants can see four different shapes of 3D objects, referents, and two virtual hands if their hand are in the field of view.}
    \label{fig:participantView}
  \end{minipage}%
  \begin{minipage}[t]{0.5\linewidth}
    \centering
    \includegraphics[width=0.9\textwidth]{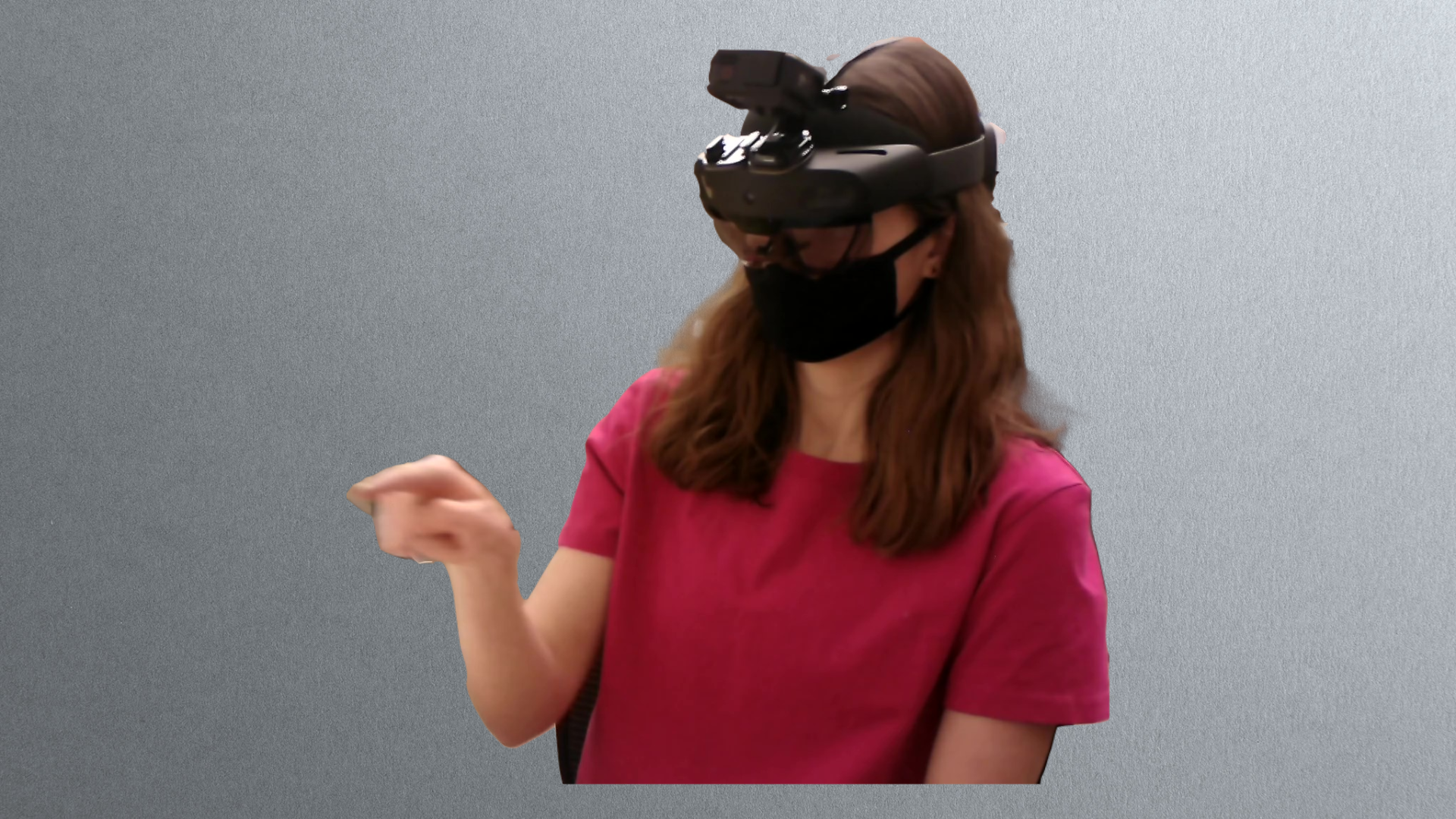}
    \caption{Experiment setup}
    \Description{A participant wears a Hololens 2 headset and a go-pro camera is mounted on top of the headset.}
    \label{fig:setup}
  \end{minipage}
\end{figure}

The experiment was conducted using Microsoft HoloLens 2 optical see-through AR head-mounted display (HMD). The system used for the experiment was developed in Unity Engine 2019.4.4f1. A GoPro Hero 7 Black was mounted on top of HoloLens 2 to record an ego-centric view of the interactions, as shown in Figure~\ref{fig:setup}. A 4k camera was placed on the front left corner facing participants to record an exo-centric view of the interactions. Two hand-shape icons on the screen were used to indicate if the hand or hands were in the view of the headset~\cite{adamVisualAid}, as shown in Figure~\ref{fig:participantView}. If either hand is out of view, the corresponding hand icon would disappear from the screen. Before starting the experiment, participants were requested to complete the informed consent and demographics questionnaire. Then participants were informed that there would be three experiment blocks with different modalities as input and they can use any interaction they feel is appropriate to execute the command based on presented text referent and input modality. Participants were told to perform gestures inside of the headset view, which they can tell by the hand icons display. The interaction modalities were presented to participants in a counter-balanced order. In each block, referents were presented in random order. The post-study questionnaire was filled out by each participant at the end of the experiment.

\subsection{Hypotheses} 
Our hypotheses were grounded in previous observations in our lab and from previous work~\cite{TVCG}: $H_1$) for the same single object manipulation referent, winning gestures in a multi-object virtual environment will be different from ones in a single object virtual environment; $H_2$) participants would prefer to use both hands for two object manipulation referents.

\section{Results}

With the experiment, 528 proposals were collected from each modality. To eliminate the effect of the referent text biasing the speech proposal~\cite{TVCG}, prior to analysis, speech proposals that were identical to the text displayed as part of the referent were removed. Resulting in 277 proposals from GS block and 261 proposals from S block.

The agreement rate ($\mathcal{AR}$), co-agreement rate ($\mathcal{CR}$) and (\textsl{V$_{rd}$}) significance test were used to determine consensus among gesture proposals~\cite{VAT+15}. $\mathcal{AR}$ is used to quantify consensus of the binned proposals for interaction by referent~\cite{AdamOrtegaElicitatioBook}, as shown in Eq.~\ref{EQ:AR}. $\mathcal{CR}$ is used to measure the amount of agreement shared between referents~\cite{VAT+15}. This study adopted Fliess's Kappa coefficient (\textsl{k$_{F}$}) and the related chance agreement term (\textsl{p$_{e}$})~\cite{TSA18} when presenting the overall agreement rate of gesture proposals. The bootstrapped 95\% confidence intervals were calculated to provide an interval estimate of each agreement score~\cite{TSA18}. We used the AGATe 2.0 tool (\underline{AG}reement \underline{A}nalysis \underline{T}oolkit)$\footnote{Available at http://depts.washington.edu/acelab/proj/dollar/agate.html}$ to assist our statistical analysis. The consensus-distinct ratio (\textsl{CDR}) was adopted to quantify the speech proposals~\cite{MOR12}. For a complete treatment on elicitation studies and methods, see Williams et al.~\cite{AdamOrtegaElicitatioBook}.
\begin{equation}
  \label{EQ:AR}
  \textsl{$\mathcal{AR}_{r}$}=\frac{\sum_{P_{i}\subset{P}}\frac{1}{2}|P_{i}|(|P_{i}|-1)}{\frac{1}{2}|P|(|P|-1)}
\end{equation}

The agreement rate $\mathcal{AR}$ for each referent r was calculated with Eq.~\ref{EQ:AR}. In Eq.~\ref{EQ:AR}, P is the set of all proposed gestures for referent r, and $P_i$ are the subsets of identical proposed gestures from P.

The overall agreement rate for gestures from G and GS blocks was .190. Based on the interpretations proposed by Vatavu and Wobbrock~\cite{VAT+15}, our study achieved a medium agreement with 12 referents and a high agreement with 4 referents. The individual agreement rate of gestures from G block and GS block alone were also calculated. The G block has .189 in agreement rate with \textsl{k$_{F}$} coefficient of .165. The chance agreement term \textsl{p$_{e}$} was .029, which indicates that the probability of agreement occurring by chance was minimal~\cite{TSA18}. The GS block obtained .193 agreement rate with \textsl{k$_{F}$} coefficient of .151, and the chance agreement term \textsl{p$_{e}$} was .050, which shows evidence of agreement beyond chance. Compared to the previous elicitation study results in the single 3D object environment~\cite{TVCG}, we have lower agreement rates in general.

\begin{figure*}
  \begin{minipage}[t]{\linewidth}
    \centering
    \includegraphics[width=0.85\textwidth]{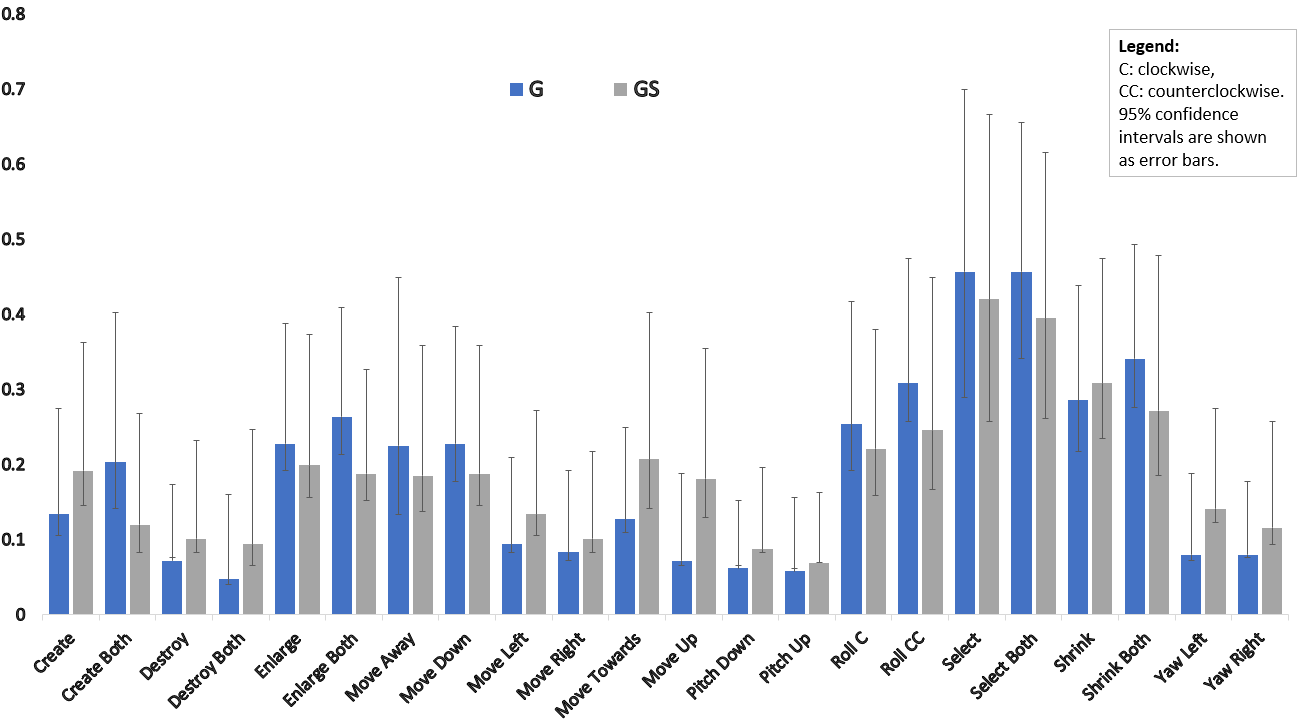}
    \caption{Gestures agreement rates in gesture only (G) block, gesture with speech (GS) block}
    \Description{A bar chart shows agreement rates by referent and modality}
    \label{fig:agreement}
  \end{minipage}%
\end{figure*}
\begin{figure*}
  \begin{minipage}[t]{\linewidth}
    \centering
    \includegraphics[width=0.85\textwidth]{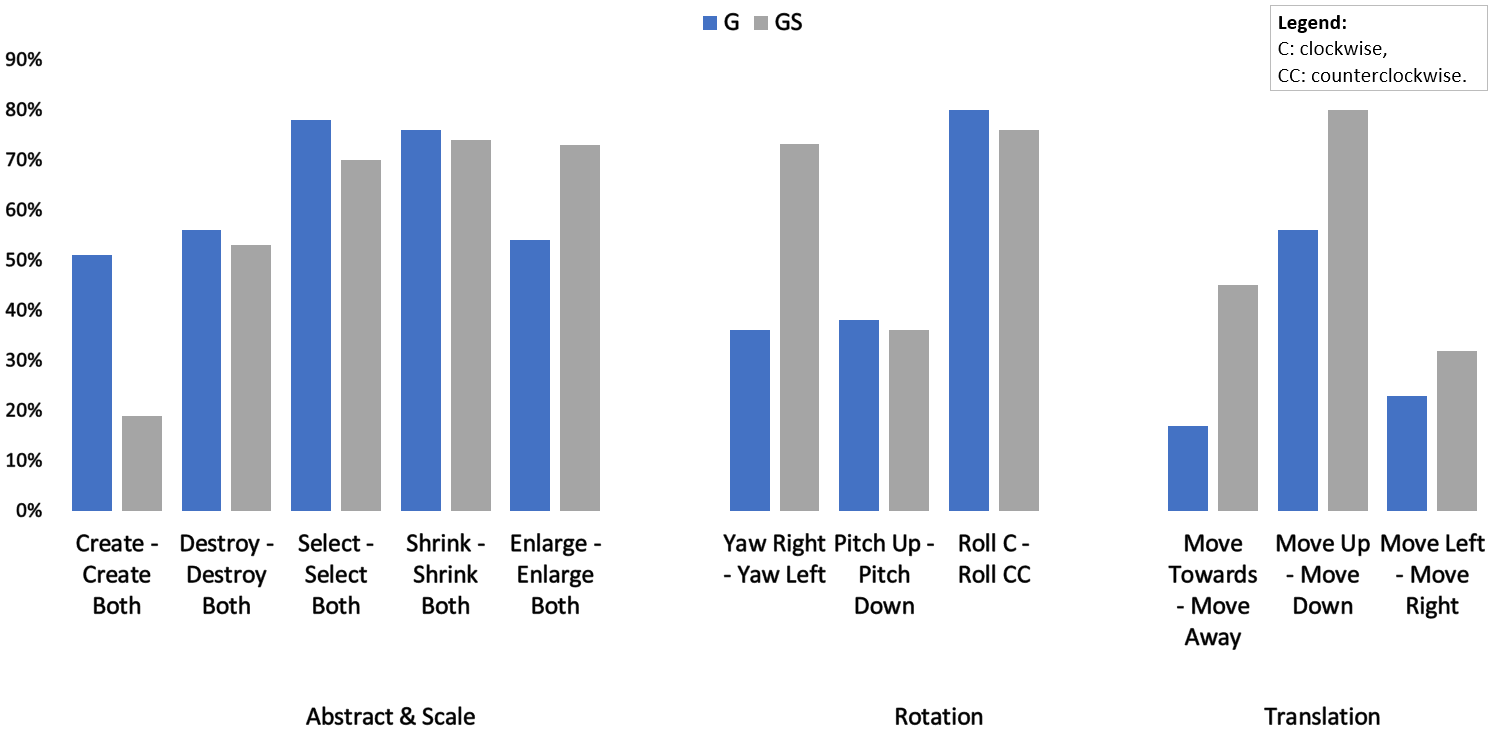}
    \caption{Gestures co-agreement between referents in gesture only (G) block and gesture with speech (GS) block}
    \Description{A bar chart shows co-agreement rates between single object and multi-object referents, and co-agreement rates between dichotomous referents.}
    \label{fig:co-agreement}
  \end{minipage}
\end{figure*}

\subsection{Unimodal Gesture and Unimodal Speech}

\subsubsection{Gesture Only}
We observed a significant effect of referent type on agreement rate in G block (\textsl{V$_{rd(21, N=528)}$} = 639.363, \textsl{p} \textless .001). The study found there were 13 referents who obtained a medium to high agreement ($\mathcal{AR} > .10$), which showed significant difference between agreement rates (\textsl{V$_{rd(12, N=312)}$} = 191.492, \textsl{p} \textless .001). Accordingly, nine referents have agreement rates below .100, which means they are in low agreement, and no significant difference in agreement rates was found (\textsl{V$_{rd(8, N=216)}$} = 7.550, \textsl{p} \textless 1.000). The highest agreement rates came from referents Select and Select Both, which are .457. The pointing gesture won the highest agreement rate for Select referent, mostly based on the natural interaction for specifying an object in the real world. The referents Shrink Both and Roll Counter Clockwise (RCC) are also achieved high agreements ($\mathcal{AR}_{ShrinkBoth} = .341$, $\mathcal{AR}_{RCC} = .308$). Among abstract referents, Destroy and Destroy Both got the two lowest agreement rates ($\mathcal{AR}_{Destroy} = .072$, $\mathcal{AR}_{DestroyBoth} = .047$). In rotation referents, Pitch up and Pitch down exhibited the two lowest agreement rates ($\mathcal{AR}_{PitchUp} = .058$, $\mathcal{AR}_{PitchDown} = .062$). For the translation referent, referent Move Up has the lowest agreement rate ($\mathcal{AR}_{MoveUp} = .072$), although Move Down has a much higher agreement rate ($\mathcal{AR}_{MoveDown} = .228$). 

A co-agreement analysis for dichotomous referents and one object versus two object referents is shown in Figure~\ref{fig:co-agreement}. The co-agreement rates of one object versus two object referents were in general higher than in dichotomous referents. The referents Select and Select Both achieved a high co-agreement ($\mathcal{AR}_{Select} = .457$, $\mathcal{AR}_{SelectBoth} = .457$, $\mathcal{CR} = .355$), which indicates 78\% of all pairs of participants have consistent gesture preference with both referents. Another high co-agreement rate came from Shrink and Shrink Both which showed 76\% of all pairs of participants that were in agreement with referent Shrink were also in agreement with gestures for referent Shrink Both ($\mathcal{AR}_{Shrink} = .286$, $\mathcal{AR}_{ShrinkBoth} = .341$, $\mathcal{CR} = .217$). 

\begin{figure}
  \begin{minipage}[t]{\linewidth}
    \centering
    \includegraphics[width=1\textwidth]{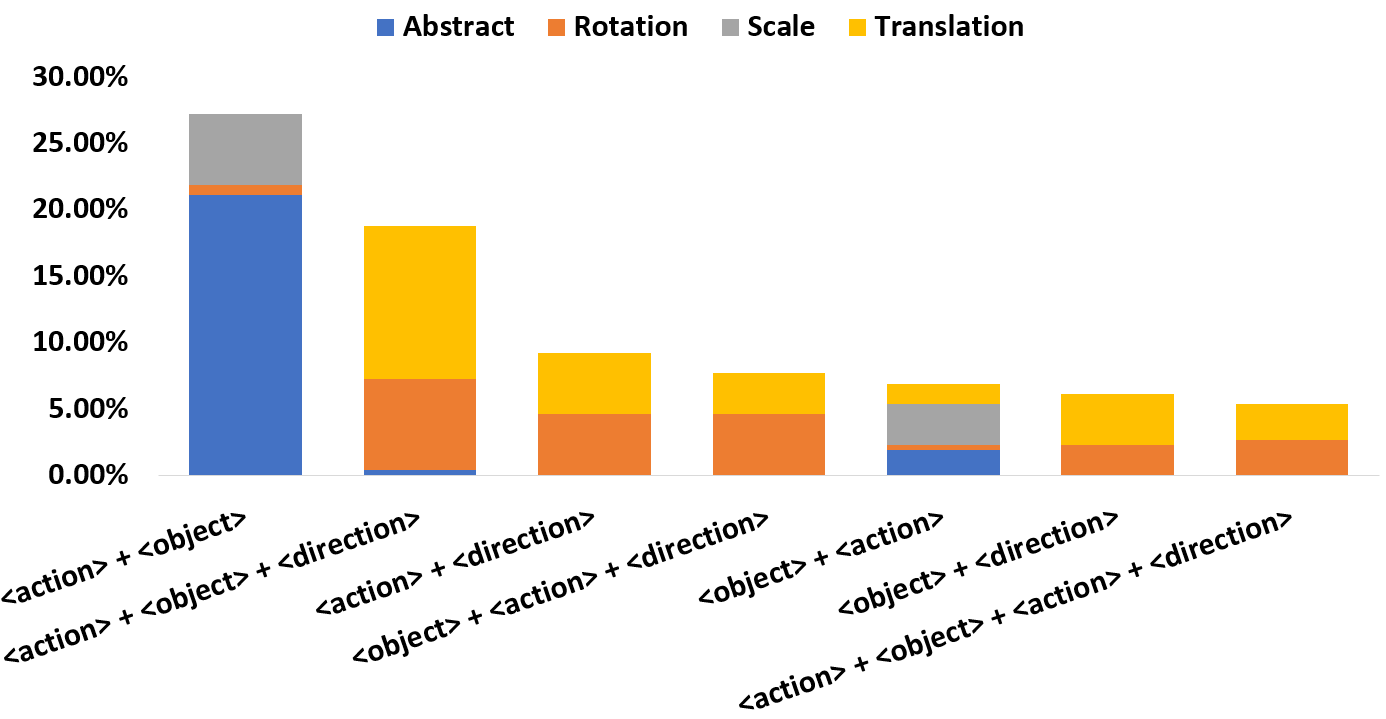}
    \caption{Usage of syntax format by referent type in the \protect\\speech only (S) block}
    \Description{A bar chart shows 80\% of syntax formats proposed in speech only block by referent category}
    \label{fig:speechOnly}
  \end{minipage}%
\end{figure}
\begin{figure}
  \begin{minipage}[t]{\linewidth}
    \centering
    \includegraphics[width=1\textwidth]{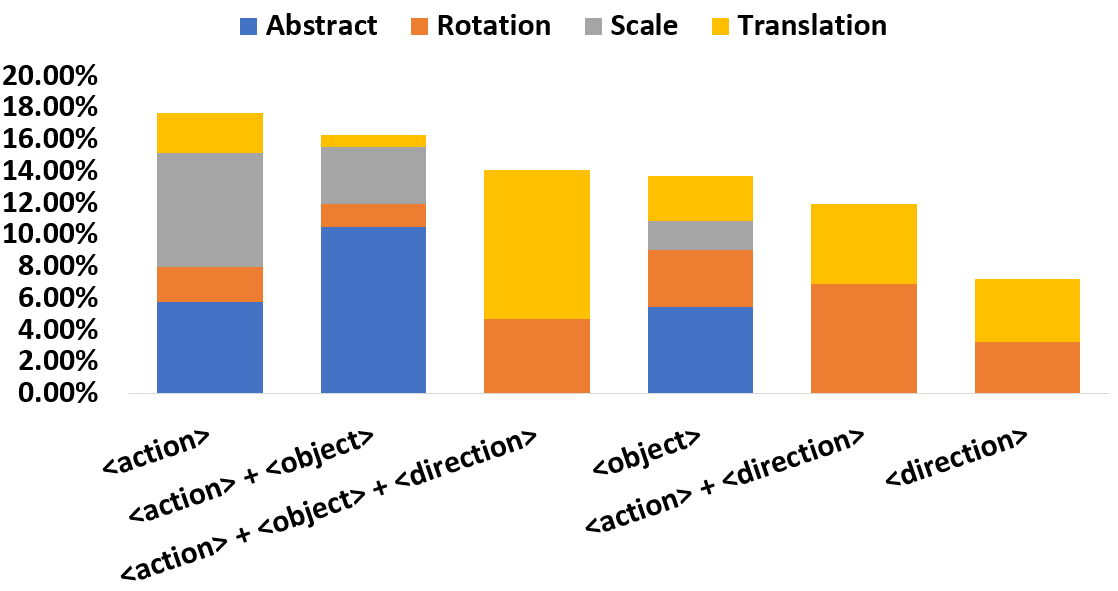}
    \caption{Usage of syntax format by referent type in the \protect\\gesture and speech (GS) block}
    \Description{A bar chart shows 80\% of syntax formats proposed in gesture and speech block by referent category}
    \label{fig:speechInGS}
  \end{minipage}
\end{figure}

\begin{table}
\footnotesize
  \centering
  \caption{Consensus-distinct ratio (\textsl{CDR}) of speech only (S) and Gesture and Speech (GS) block by referent category}
  \begin{tabular}{lp{2cm}p{2cm}p{2cm}}
    \toprule
     Referent Category & Speech Only & Gesture and Speech\\
    \midrule
    Abstract & 43.75\% & 35.21\% \\
    Rotation & 24.56\% & 16.44\% \\
    Scale & 57.14\% & 32.0\% \\
    Translation & 42.65\% & 22.89\% \\
    \bottomrule
  \end{tabular}
  \label{table:CDR}
\end{table}

\subsubsection{Speech Only} 
For speech data, we adopted the binning criterion wherein ``enlarge yellow and green'' and ``enlarge cube and sphere'' were equal to ``enlarge yellow cube and green sphere''. However, ``yellow cube pitch down'' and ``yellow cube rotate down'' were counted as different proposals. 

Table~\ref{table:CDR} shows the consensus-distinct ratio (\textsl{CDR}) of different categories of referents in the S block. The \textsl{CDR} is used to calculate the percent of distinct speech proposals by referent that achieved a consensus threshold of two~\cite{MOR12}. The results demonstrated that scale referents have the highest \textsl{CDR}, in addition to abstract referents which present a \textsl{CDR} that are almost twice high when compared to 24.52\% from the previous elicitation study with a single 3D object~\cite{TVCG}. The rotation referents hold the lowest \textsl{CDR}. Based on the data, a low \textsl{CDR} could be caused by different expressions of rotation. For example, ``spin'' or ``rotate'' plus gesturing direction was proposed to achieve ``roll'', ``yaw'', or ``pitch''. A similar finding was presented in the previous elicitation study with a single 3D object~\cite{TVCG}. There are few alternative phrases for ``move up / down / left / right, '' which could explain that translation referents have a higher \textsl{CDR} than rotation referents. Similarly, less options of replacement for action or status phrases such as ``shrink'' and ``smaller'' in scale proposals. Figure \ref{fig:speechOnly} presents the syntax formats covered more than 80\% of proposals in the S block. It is obvious that $\langle action \rangle$ $\langle object \rangle$ and $\langle action \rangle$ $\langle object \rangle$ $\langle direction \rangle$ are the most common formats for speech proposals. Moreover, rotation and translation referents elicited more variants of syntax, which means that various syntax should be considered while designing unimodal speech commands.

Despite bias from text referents, participants often preferred interaction from left to right with multiple 3D objects. For example, with the referents of ``create two objects at the same time'', two participants proposed ``create green sphere and yellow cube'', even though all text referents involving two objects started as ``yellow cube and green sphere'' in the experiment. It shows that participants favored creating objects starting from the left since the green sphere was placed to the left side of the yellow cube in the scene.

\subsection{Multimodal interaction: Speech and Gesture}
\subsubsection{Gesture in GS}
The results additionally demonstrate that the referent type has significant effect on gesture agreement rates in GS block (\textsl{V$_{rd(21, N=528)}$} = 361.624, \textsl{p} \textless .001). There were 19 referents who achieved medium to high agreement ($\mathcal{AR} > 0.10$), and presented significant difference between agreement rates (\textsl{V$_{rd(18, N=456)}$} = 262.325, \textsl{p} \textless .001). Only 3 referents have low agreement rates ($\mathcal{AR}_{DestroyBoth} = 0.094$, $\mathcal{AR}_{PitchUp} = 0.069$, $\mathcal{AR}_{PitchDown} = 0.087$), and further significant differences among those agreement rates were not found (\textsl{V$_{rd(2,N=72)}$} = 1.368, \textsl{p} \textless 1.000). The highest agreement rate in the GS block was from referent Select ($\mathcal{AR}_{Select} = 0.42$), and referent Select Both, who was not far behind in rank. ($\mathcal{AR}_{SelectBoth} = 0.395$). As in the G block, referent Shrink also obtained a high agreement rate while combining with speech ($\mathcal{AR}_{Shrink} = 0.308$). Moreover, referents Destroy and Destroy Both showed a similar low agreement as in the G block, compared to other referents ($\mathcal{AR}_{Destroy} = 0.101$, $\mathcal{AR}_{DestroyBoth} = 0.094$). As shown in Figure~\ref{fig:agreement}, the two lowest agreement rates in the GS block came from referents Pitch Up and Pitch Down.

In terms of co-agreement, for one object versus two object referents, the average co-agreement rate was 68\% without including referent Create Both. This finding indicates that that a high number of participants kept the same preferences for both one object and two object manipulation. The cause of a low co-agreement rate between referent Create and Create Both could be the low agreement rate for Create Both in the GS block ($\mathcal{AR}_{Create} = .192$, $\mathcal{AR}_{CreateBoth} = .120$, $\mathcal{CR} = .036$). Higher co-agreement rates were found for dichotomous referents compared to the G block. As shown in Figure~\ref{fig:co-agreement}, the co-agreement rates of translation referents were increased the most compared to the values in the G block, which indicates multimodal interaction assisted participants achieving more agreement for dichotomous translation referents.

\subsubsection{Speech in GS} 
Figure \ref{fig:speechInGS} shows syntax formats covered more than 80\% of proposals in the GS block. Compared to the S block, participants have proposed a fair amount of single-word commands with compensation from gestures. These single-word proposals included $\langle action \rangle$ only, $\langle object \rangle$ only, $\langle direction \rangle$ only, and $\langle status \rangle$ only. All proposals with single-word commands account for 39.71\% of the total proposals in the GS block. In contrast, the proportion of single-word proposals in the S block were merely 6.48\%. The prior study mentioned that part of the $\langle action \rangle$ $\langle object \rangle$ syntax proposed in the S block turned into $\langle action \rangle$ plus gesture proposals in the GS block~\cite{TVCG}. In the GS block, the most used syntax format was $\langle action \rangle$ only, shown in all four categories of referents. If the speech does not indicate the target, it can then be assumed that gestures were used for identifying the target object in a multi-object virtual environment. As anticipated, based on the results, 87.5\% of proposals with $\langle action \rangle$ only syntax format have involved gestures of``pointing'', ``tapping'', or ``grabbing''. Furthermore, with $\langle direction \rangle$ only syntax proposals, 84.21\% of gestures showed ``pointing'' or ``tapping'' to indicate the target object. In contrast, proposals consisting of the $\langle object \rangle$ only syntax format had merely 28.95\% of the proposals involving the gestures ``pointing'' or ``grabbing''. This result proved the complementary feature of multimodal interaction. Due to the necessity of identifying the target object in a multi-object environment for manipulation and the flexibility of using speech that multimodal interaction gave participants, less agreement was shown with speech proposals in the GS block compared to in the S block (Table~\ref{table:CDR}). 

\subsubsection{Gesture and Speech Association}
The study looked into the association between the stroke of a gesture proposal and the corresponding speech proposal in the GS block. A stroke is considered the peak of effort for a specific gesture~\cite{MCN05}, which holds the meaningful content of the gesture. We classified the main speech content into three types of expressions (nominal, deictic, verb) based on prior work from Bourguet and Ando~\cite{BOU+98}. During the video annotation, recordings were made of the expressions in relation to speech content while the main stroke of the gesture occurred. The study found that strokes for abstract referents were mainly associated with nominal expressions, such as ``the yellow cube'' or ``objects''. The referents Destroy and Destroy Both were exceptions, which could be related to the low agreement on gesture proposals. All scale strokes were more synchronized with verb expressions, mostly ``enlarge'' and ``shrink''. It should be noted that there were much fewer deictic expressions used in the scale speech proposals, which indicates the limitation of associated expressions. In terms of the translation and rotation referents, 9 out of 12 showed a strong association between strokes and deictic expressions. For example, participants would execute the stoke of pitch up while saying ``up''. The Move Away and Yaw Right referents were slightly more synchronized with verb expressions, and the stroke of roll counterclockwise showed more association with nominal expressions. 

\begin{figure*}
      \includegraphics[width=1\textwidth]{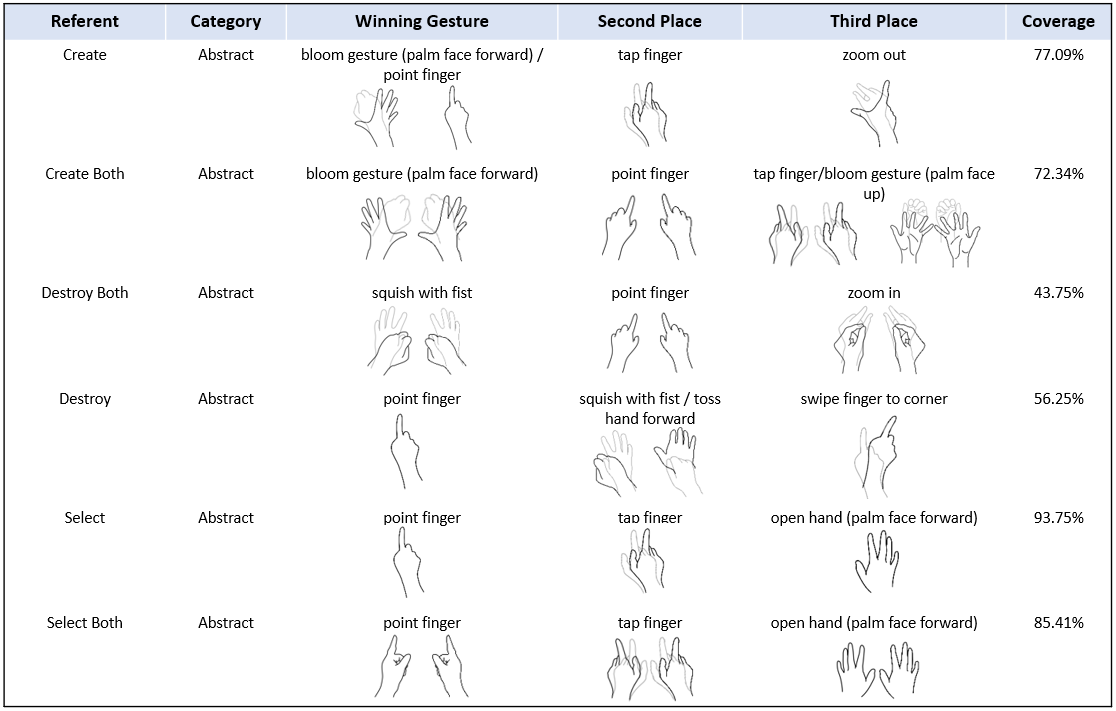}
    \caption{Top three proposed gesture variants for abstract referent}
    \Description{Gesture sketches based on top three proposed gesture variants for abstract referent}
    \label{fig:abstractgestures}
\end{figure*}

\begin{figure*}
      \includegraphics[width=1\textwidth]{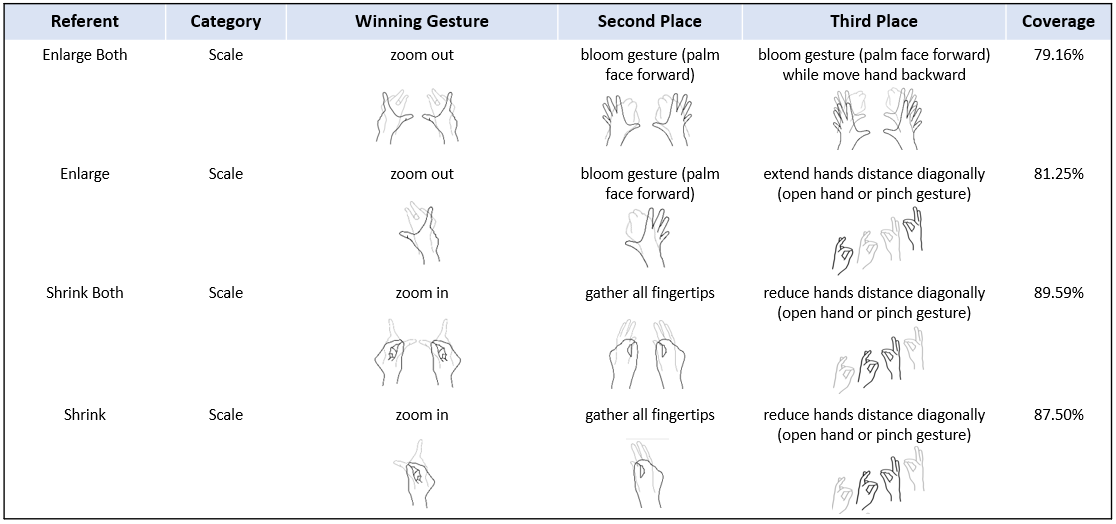}
    \caption{Top three proposed gesture variants for scale referent}
    \Description{Gesture sketches based on top three proposed gesture variants for scale referent}
    \label{fig:scalegestures}
\end{figure*}

\begin{figure*}
      \includegraphics[width=1\textwidth]{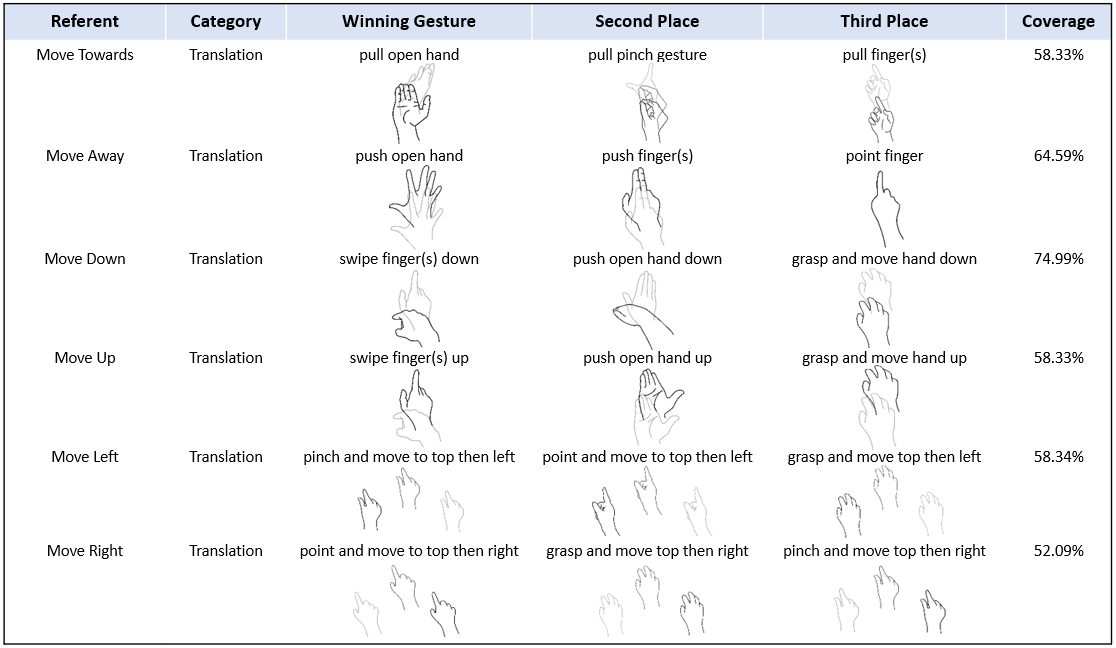}
    \caption{Top three proposed gesture variants for translation referent}
    \Description{Gesture sketches based on top three proposed gesture variants for translation referent}
    \label{fig:translationgestures}
\end{figure*}

\begin{figure*}
      \includegraphics[width=1\textwidth]{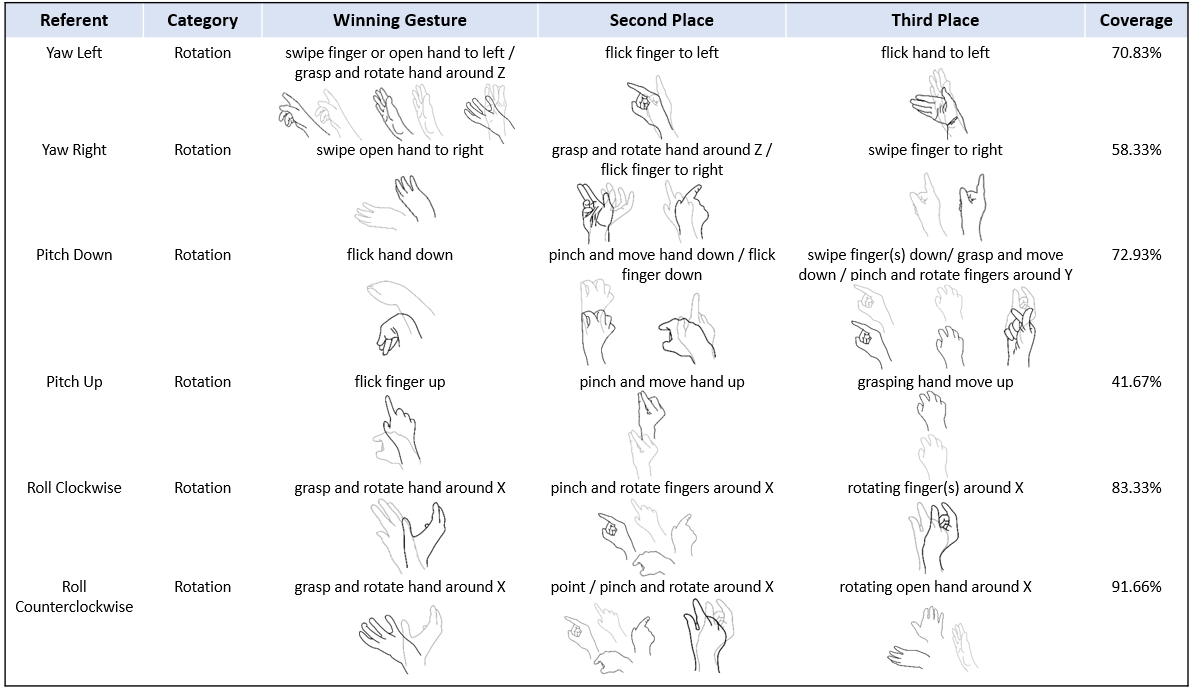}
    \caption{Top three proposed gesture variants for rotation referent}
    \Description{Gesture sketches based on top three proposed gesture variants for rotation referent}
    \label{fig:rotationgestures}
\end{figure*}

\section{Discussion and Design Guidelines}
The results support our hypothesis $H_2$ that the consensus set of gestures indicates that participants preferred to use both hands for two virtual object manipulation. Chi-square analysis showed that the difference between
one hand and two hands adoption in the two virtual environments was statistically significant ($X^2$ = 255.33, \textsl{p} \textless .001). This means multi-object environment increases the usage of both hands. The results support $H_1$ for some tasks, because there were 11 out of 17 single object manipulation referents resulting with different winning gestures, compared to ones from a single object virtual environment. In general, there are similarity and dissimilarity in multimodal interaction with an multi-object virtual environment and single object virtual environment. The multi-object environment has inspired more physical interaction which came from experience with real world. 

Based on our resulting top three proposed variants of each referent, among 17 single object manipulation tasks, six referents have the same winning gestures as prior findings in a single object virtual environment~\cite{TVCG, PHA+18}. Another five referents' second place proposals were identical to previous results in a single object virtual environment~\cite{TVCG, PHA+18}. Within the six referents that have the same winning gestures in both virtual environments, two of them are Shrink and Enlarge from scale referents, three translation referents are Move Toward, Move Away, and Move Left, and one rotation referent is Yaw Left. Legacy bias is an issue in elicitation study that uses' gesture proposals are biased due to the previous experience with exist interfaces~\cite{MOR+14}. The legacy bias from interaction with a multi-touch screen could contribute to the identical scaling proposals. The ``screwing in a light bulb'' gesture for rotating around the Z axis was also found in Williams et al., and Pham et al.'s works~\cite{TVCG, PHA+18}. Unsurprisingly, abstract referents have less similarity on their proposals. The winning gesture for creating an object in this work used gathered and then spread fingertips as the original blooming gesture from HoloLens 1~\cite{hl1}, but with the palm facing forward instead of facing up. Due to the difference, we do not consider that our blooming gesture came from legacy bias and more likely was a spontaneous proposal that could inspire future gesture designing for AR interaction. 

In terms of speech proposals, our results showed more variety in syntax formats. We have two more single-word syntax formats in GS block compared to ones found in Williams et al.~\cite{ISS}, and they were $\langle object \rangle$ only and $\langle status \rangle$ only. For speech only interaction, our study presented $\langle action \rangle$ + $\langle object \rangle$ as the top rank syntax format, compared to the $\langle action \rangle$ only syntax format which has a similar proportion in a single object environment~\cite{ISS}. We believe this result was due to the multiple object environment in our study, and participants tended to specify the target object for interaction. 

The results of the study found that participants preferred symmetric bimanual versions of the single-handed gesture for two object manipulation. For example, the winning proposal for shrinking a single object was the zoom in gesture, and the winning gesture for shrinking two objects side by side was to perform zoom in with both hands simultaneously. This result of symmetric bimanual interaction is reasonable since both targets were inside the participant's field of view, which made symmetric action easy to perform~\cite{BAL+00}. The exception of destroying proposals could be related to the low agreement rate for both destroy referents, which indicates people have less common sense for destroying from reality-based interaction~\cite{JAC+08}. According to the answers in the post-study questionnaire, 13 out of 24 participants expressed that it was fairly natural to think of using both hands for two object manipulation. Five participants indicated it was harder to develop the proposal for two object interaction compared to the single object manipulation. One participant said that the single hand gesture could be used to replace two hand interaction as needed. Our findings could be used to develop gesture recognizers for a multi-object virtual environment by sensing the user's intent based on the hands involved.

Speech recognition with an AR headset is difficult due to the environment noise, unintended commands, and sometimes the accent of the user. With the knowledge of the association between speech expressions and gesture stroke, a more specific hypothesis can be implemented in the recognition system to improve speech detection efficiency and accuracy in AR. While the previous study only focused on pointing gestures~\cite{BOU+98}, our study discovered the association between common manipulation gestures and speech commands for interaction in a multi-object virtual environment. 

\textbf{Design Guidelines --} Based on the user-defined gesture sets from our study and literature, while some gestures and speech syntax formats remain similar, there were differences in multimodal interaction between a single object and a multi-object virtual environments. Participants' proposals in our study showed more physical interactions such as pinching or grasping the target object and ``turning a doorknob'' for rotation tasks. Similar to prior findings suggested to include aliasing for gestures and speech~\cite{WOB+09, MOR12, TVCG}, we propose that including aliasing could significantly improve the performance of the recognizer. For example, using the commands ``spin'' or ``rotate'' plus gesture indicates direction should be equal to use commands ``roll/yaw/pitch''. With gestures, performing pinching or grasping then moving the hand for virtual object translation should be equivalent to pointing at the target then moving the finger. Our results indicate that implementing the top three proposed variants (Figure~\ref{fig:abstractgestures}, Figure~\ref{fig:scalegestures}, Figure~\ref{fig:translationgestures}, Figure~\ref{fig:rotationgestures}) of a gesture could increase the coverage of proposed gestures to 70\% on average. The variety of syntax formats in the GS block indicates that various combinations of speech and gesture could be designed for interaction in an augmented reality environment. Moreover, as Williams and Ortega mentioned in their work, legacy bias could be a benefit to new technology because it is memorable and discoverable~\cite{adamLegacy}. We suggest that emerging technology such as AR-HMD should consider both legacy bias from the touchscreen and physical interaction based on body awareness and environmental skills~\cite{JAC+08}.

\section{Limitation and Future Work}
The text referents could bias participants' speech proposals in our experiment. We also know that using animation as referents would bias the gesture proposal in the elicitation study ~\cite{TVCG, KHA+19}. It is still a research question that how to eliminate the bias from referent presentation. Our experiment design requires participants to give both speech and gesture in GS block, which could end with the unnatural speech proposals from participants. Therefore, we will use a more efficient but flexible way to elicit proposals from participants in our future elicitation study. For example, we could adopt the "before" and "after" approach to present the desired effect of a referent for our future study ~\cite{PHA+18, SEY+12}. Reducing the fatigue caused by mid-air interactions is another necessary vein of future work. One way to mitigate this issue is to use other modalities such as eye-gazing combined with speech to replace mid-air gestures. Another option for reducing fatigue could be developing microgestures that require less psychical effort than mid-air gestures. 

\section{Conclusion}
This study investigated multimodal interaction in a multi-object virtual environment. We chose 22 referents for the elicitation study that included canonical referents for scale, translation, and rotation tasks and three abstract referents. We generated a consensus set of gestures for interaction in a multi-object virtual environment and found that participants used the same gesture for one and two objects but with both hands for two object manipulation. The results further demonstrated that participants tended to act on the target objects in a multi-object virtual environment, indicating more physical interaction where preferred. Further, in the study, more speech syntax formats were proposed in multimodal interaction in a multi-object virtual environment. We discovered the association between expressions and stroke, which can improve the accuracy and efficiency of the recognition system. We also provided design guidelines based on our findings and comparison with prior works in a simple virtual environment. 

\begin{acks}
To Robert, for the bagels and explaining CMYK and color spaces.
\end{acks}

\bibliographystyle{ACM-Reference-Format}
\bibliography{elicitation-papers}

\appendix

\end{document}